\documentstyle[aps,12pt]{revtex}
\begin{document}
\title{Comment on ''Orbitally degenerate spin-1 model for insulating V$_{2}$O%
$_{3}$'' }
\author{R.J. Radwanski }
\address{Center for Solid State Physics, S$^{nt}$ Filip 5, 31-150 Krak\'{o}%
w, Poland,\\
Institute of Physics, Pedagogical University, 30-084 Krak\'{o}w, Poland.}
\author{Z. Ropka}
\address{Center for Solid State Physics, S$^{nt}$ Filip 5, 31-150 Krak\'{o}%
w, Poland.\\
e-mail: sfradwan@cyf-kr.edu.pl}
\maketitle

\begin{abstract}
We criticize the model proposed by Mila et al. (Phys.Rev.Lett. 85 (2000)
1714) for the description of the V$^{3+}$ ion in V$_{2}$O$_{3}$ by pointing
out that it is completely artificial because the condition about the
degenerate (doublet) orbital ground state required for the authors' model
cannot be realized in the reality. We claim that the V$^{3+}$ ion should be
considered as described by the quantum numbers S=1 and L=3 and with taking
into account the orbital moment and the intra-atomic spin-orbit coupling. \ 
\end{abstract}

\date{(5.12.2000)}

In a recent Letter [1] Mila et al. have proposed a new ''orbitally
degenerate spin-1 model for the insulating V$_{2}$O$_{3}$''. By this Comment
we criticize [2] this model pointing out that the model proposed by authors
for the description of the V$^{3+}$ ion in V$_{2}$O$_{3}$ is completely
artificial because i) the condition about the degenerate (doublet) orbital
ground state required for the authors' model cannot be realized in the
reality. Moreover, we criticize ii) the description of the V$^{3+}$ ion in V$%
_{2}$O$_{3}$ by an ''orbitally degenerate spin-1 model'' (i.e. with S=1 and
L=1/2) as we iii) claim that the V$^{3+}$ ion in V$_{2}$O$_{3}$ should be
described by the quantum numbers S=1 and L=3 [3,4].

The critiques i) and ii) are associated with the Jahn-Teller theorem that
states that each system tends to, if possible, to the singlet ground state.
Thus, the orbital ground state, demanded by the authors' model, will be not
realized in the reality.

The claim iii) follows from the two Hund's rules (two rules, i.e. 1$^{o}$
the maximal S and 2$^{o}$ the maximal L for two d electrons) [5]. These two
Hund's rules yield S=1 and L=3. The 21-fold degeneracy will be removed by
crystal-field and spin-orbit interactions [3,4], and finally by
spin-dependent interactions yielding a singlet ground state of the V$^{3+}$
ion in the atomic scale. Despite of the singlet ground state state magnetism
can develop by moment-induced mechanism like in many praseodymium compounds.
Our understanding of 3d-ion compounds is close to an original idea of Van
Vleck from 1932 [6], that electronic and magnetic properties are largely
determined by the atomic-like electronic structure.

In favour of the authors' model we would like to point out that their model
with S=1 and L=1/2 is already closer to the reality than a model of \
Castellani et al. [7], a S=1/2 orbitally degenerate model, and a S=1 model
without an orbital degeneracy [8]. These different models, currently
discussed, clearly show that a description of V$_{2}$O$_{3}$, as well as
many other 3d-ion compounds, is far from being established calling for the
normal open scientific discussion.

In a conclusion, the model proposed by authors is much artificial because
the condition about the degenerate (doublet) orbital ground state required
for the authors' model cannot be realized in the reality. We claim that the V%
$^{3+}$ ion in V$_{2}$O$_{3}$ should be considered as described by the
quantum numbers S=1 and L=3. We are convinced that taking into account the
orbital moment and the intra-atomic spin-orbit coupling is indispensable for
the physically adequate description of electronic and magnetic properties of
V$_{2}$O$_{3}$.

E-mail for correspondence: sfradwan@cyf-kr.edu.pl

\end{document}